\documentclass[prb,twocolumn,showpacs,amsmath,amssymb,floatfix]{revtex4}
\usepackage{graphicx}
\usepackage{dcolumn} 
\usepackage{bm}
\usepackage{amsmath}
\usepackage{latexsym}
\usepackage{amsfonts}
\usepackage{dcolumn}
\usepackage{amssymb}
\usepackage[rightcaption]{sidecap}
\newcommand{\be}{\begin{equation}}
\newcommand{\ee}{\end{equation}}
\newcommand{\bea}{\begin{eqnarray}}
\newcommand{\eea}{\end{eqnarray}}
\newcommand{\nn} {\nonumber}

\newcommand{\Tr}{ {\rm Tr} \, }

\def\ve{\varepsilon}

\def\l{\lambda}
\def\L{\Lambda}

\def\S{\Sigma}

\def\vf{\varphi}

\def\w{\omega}

\def\inf{\infty}
\def\ra{\rightarrow}
\def\bra{\langle}
\def\ket{\rangle}

\def\xc{{\rm xc}}

\def\Tr{{\rm Tr}\,}

\def\br{\mbox{\boldmath $r$}}

\begin{document}
\title{Linear density response function within the time-dependent\\ exact-exchange approximation}
\date{\today}
\author{Maria Hellgren}
\author{Ulf von Barth}
\affiliation{Mathematical Physics, Institute of Physics, Lund University, 
S\"olvegatan 14A, S-22362 Lund, Sweden}  
\date{\today}              
\begin{abstract}
We have calculated the frequency-dependent exact exchange (EXX) kernel of
time-dependent (TD) density functional theory employing our recently proposed
computational method based on cubic splines. With this kernel we have calculated
the linear density response function and obtained static polarizabilites, 
van der Waals coefficients and correlation energies for all spherical spin 
compensated atoms up to Argon. Some discrete excitation energies have
also been calculated for Be and Ne. As might be expected, the results of the
TDEXX approximation are close to those of TD Hartree-Fock theory. In addition,
correlation energies obtained by integrating over the strength of the Coulomb
interaction turn out to be highly accurate. 
\end{abstract}
\maketitle
\section{Introduction}
The present paper is one in a series of papers\cite{abl,tddftvar,kvB,hvB} reporting on 
work with the overall aim of finding computationally efficient but still accurate ways of
calculating excited-state properties of systems in which excitonic effects play an important 
role. Traditionally, such effects have been studied by solving approximate versions of the 
Bethe-Salpeter equation in which the particle-hole interaction is taken to be a statically screened 
Coulomb potential.\cite{bsenote} Unfortunately, such ab initio methods are computationally very demanding especially in low symmetry systems such as nano structures and large molecules.
During the last decade, time-dependent density functional theory\cite{rg0,pggtrue,revlucia} (TDDFT) 
has emerged as a competing technique due to its computational efficiency 
and better scaling with the size of the system. In recent years, the use of TDDFT has virtually 
'exploded' within theoretical chemistry. On the other hand, within TDDFT,  the 
limitations are rather caused by our rudimentary knowledge of
the 'mysterious' exchange-correlation (XC) kernel $f_\xc$, into which
all effects beyond the RPA (Random Phase Approximation) are 
transferred. 

In most applications of TDDFT one resorts to an XC kernel constructed from some
ground-state XC potential evaluated at the instantaneous electron density. In this way, the 
non-locality in time (memory effects) leading to a frequency-dependent XC kernel, 
is neglected. These are the so-called adiabatic approximations and the simplest 
example is the adiabatic local density approximation (ALDA) derived from the 
ground-state local density approximation. In fact, any approximate functional within 
ground-state density functional theory (DFT) can yield an adiabatic approximation within TDDFT. Although such approximations have been shown to provide good estimates of many physical quantities, 
there are qualitative features which cannot be accounted for. For the description of, e.g., excitation 
energies with multiple-particle character the kernel is expected to have a strong frequency-dependence.\cite{m1}

Alternative approaches based on many body perturbation theory (MBPT) to generate new and more advanced kernels have been proposed by some authors.\cite{ro,mr,tp.2001,lucia,tddftvar} Up to now, however, the performance of such kernels in describing excited- as well as ground-state properties has been the subject of little investigation, especially in finite systems.  

The simplest approximation derived from MBPT is the time-dependent
exact-exchange (TDEXX) approximation. It can be obtained from a stationary action 
principle by retaining only the exchange part of the XC part of the total
action, i.e., terms up to first order in the Coulomb interaction. 
The exact-exchange (EXX) kernel  $f_{\rm x}$ is frequency-dependent and therefore fundamentally differs from
the adiabatic approximations. Because the EXX kernel can also be derived from our
\cite{tddftvar} variational approach to the many-body problem, as is done here, it automatically posesses conserving properties
which means, e.g., that the resulting linear density response function will
obey the $f$-sum rule. Furthermore, the corresponding EXX potential of ground-state
DFT has already been thoroughly studied and shown to share many properties of the
exact XC potential like, e.g., the correct $-1/r$ asymptotic decay and the
derivative discontinuity with respect to particle number. 

Implementations of the TDEXX approximation has so far been limited to the
calculation of the total energy and the plasmon dispersion relation of the
electron gas,\cite{kvB} the optical absorption spectrum of bulk silicon,\cite{kg2}
and, recently, van der Waals coefficients and polarizabilities for some
simple atoms and molecules.\cite{shh2006} The adiabatic TDEXX and the exact TDEXX
of a two-electron system (which turns out to be frequency-independent)
known as the PGG (Petersilka, Gossman, and Gross) approximation have also been used
in the calculation of atomic and molecular transition 
frequencies.\cite{higb2002,pgb,gpg} In the non-linear regime the TDEXX 
approximation has most recently been applied to the problem of electron dynamics
in a quantum well. \cite{Wiull2008}

In this paper, we calculate the linear density response function using the {\em fully}
frequency-dependent EXX kernel for all spherical spin compensated atoms up to Argon
and present results on correlation energies, van der Waals coefficients, static
polarizabilites and a few discrete excitation energies for Beryllium and Neon.
The correlation energies, calculated from the Hellman-Feynman theorem applied to the
strength of the Coulomb interaction, turn out to be very accurate, whereas
polarizabilities and van der Waals coefficients are similar in quality to the rather poor
results of time-dependent Hartree-Fock theory (TDHF).  

The paper is organized as follows. In Sec. \ref{theory} we sketch the derivation of the
relevant equations and discuss the TDEXX approximation in comparison to TDHF. 
We also give a short description of the computational methods. 
In Sec. \ref{result} we present our results and compare them with other approximations and exact results. Some attention is given to the kernel 
itself and we provide evidence of the  $f$-sum rule being obeyed by studying the large 
$\w$-behavior of both $f_{\rm x}$ and the dynamical polarizability.  

As mentioned above, the fact that the TDEXX approximation obeys the $f$-sum rule
follows from the possibility to derive it from the variational and conserving approach
to MBPT. In the Appendix we show, however, the details on how the defining 
equations lead to the $f$-sum rule. This is rather illuminating and 
demonstrates the necessity for using the correct local exchange potential from
the LSS equation in the evaluation of the $f_{\rm x}$ kernel in order to have
the sum-rule fullfilled.

Finally, in Sec. \ref{conclu}, we draw our conclusions and announce a forthcoming 
publication on spectral properties. 
\section{Theory and computational details}
\label{theory}
\subsection{General formulation}
Within TDDFT the electronic linear density response function $\chi$ is given by 
\be
\chi=\chi_s+\chi_s(v+f_\xc)\chi,
\label{rpafxc}
\ee
where $\chi_s$ is the Kohn-Sham (KS) linear density response function, $v$ is the Coulomb interaction 
and $f_\xc$ is the XC kernel defined as the functional derivative of the XC potential $v_\xc$,
\be
f_\xc=\frac{\delta v_\xc}{\delta n}.
\ee

It has been shown in previous publications,\cite{tddftvar,hvB} that various consistent and, in particular, conserving approximations to $v_\xc$ and $f_\xc$  can be obtained from the Klein action functional\cite{klein}  by choosing physically reasonable approximations to the defining $\Phi$-functional.\cite{baym} The stationary property of the Klein functional with respect to Green functions generated by local potentials leads directly to the linearized Sham Schl\"uter\cite{ss} (LSS) equation 
\be
\int \chi_s(1,2)v_{\xc}(2)d2=\int \S_s(2,3)\Lambda(3,2;1)d2d3,
\label{lsseq}
\ee
where the self energy $\S_s$ is $\Phi$-derivable and expressible in only the KS Green function $G_s$ and the Coulomb interaction. The quantity $\L$ is the functional derivative of  $G_s$ with respect to the KS potential $V$,
$$
i\Lambda(3,2;1)=\frac{\delta G_s(3,2)}{\delta V(1)}=G_s(3,1)G_s(1,2).
$$
The equation for the corresponding kernel $f_{\xc}$ is obtained by varying Eq.~(\ref{lsseq}) with respect to $V$. The result is:
\begin{eqnarray}
&&\int \chi_s(1,2)f_{\xc}(2,3)\chi_s(3,4)d2d3\nn\\
&&\,\,\,\,\,\,\,\,\,\,=\int \frac{\delta\S_s(2,3)}{\delta V(4)}\Lambda(3,2;1)d2d3\nn\\
&&\,\,\,\,\,\,\,\,\,\,\,\,\,\,+\int \Lambda(1,2;4)\Delta(2,3)G_s(3,1)d2d3\nn\\
&&\,\,\,\,\,\,\,\,\,\,\,\,\,\,+\int G_s(1,2)\Delta(2,3)\Lambda(3,1;4)d2d3,
\label{fxceq}
\end{eqnarray}
where $\Delta(2,3)=\S_s (2,3)-v_{{\xc}}(2)\delta(2,3)$. Due to the variational property of the Klein functional and the $\Phi$-derivability of the self energy the kernel $f_\xc$ obtained from Eq.~(\ref{fxceq}) will result in a response function (Eq.~(\ref{rpafxc})) which is particle conserving. In the linear regime this means, e.g., that it must obey the $f$-sum rule. 
\begin{figure}
\includegraphics[width=8.5cm, clip=true]{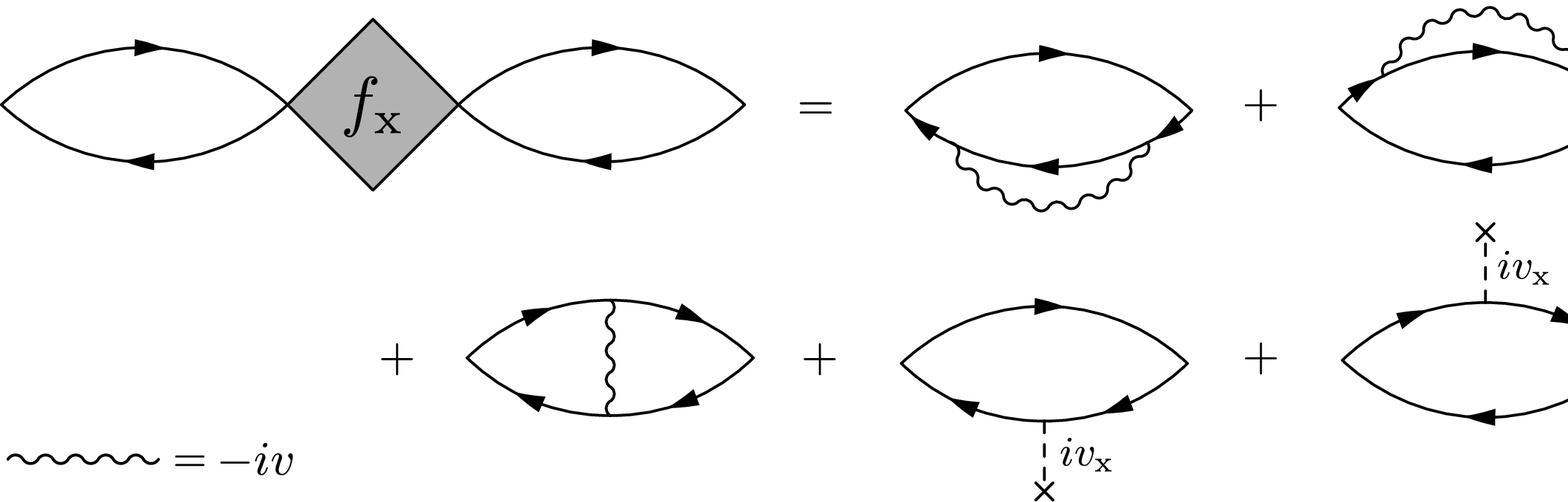}\\
\caption{Diagrammatic representation of Eq. (\ref{fxceq}) with the self energy in the Hartree-Fock approximation. }
\label{diagram}
\end{figure}

In this work we are interested in studying the so-called TDEXX approximation which is derived at the TDHF level of MBPT. The self energy is then given by 
\be
\S^{\rm x}_s(2,3)= iv(2,3)G_s(2,3),
\ee
and its variation with respect to $V$ becomes 
\be
\frac{\delta \S^{\rm x}_s(2,3)}{\delta V(4)}=-v(2,3)\Lambda(2,3;4).
\ee
\begin{figure*}
\includegraphics[width=17.5cm, clip=true]{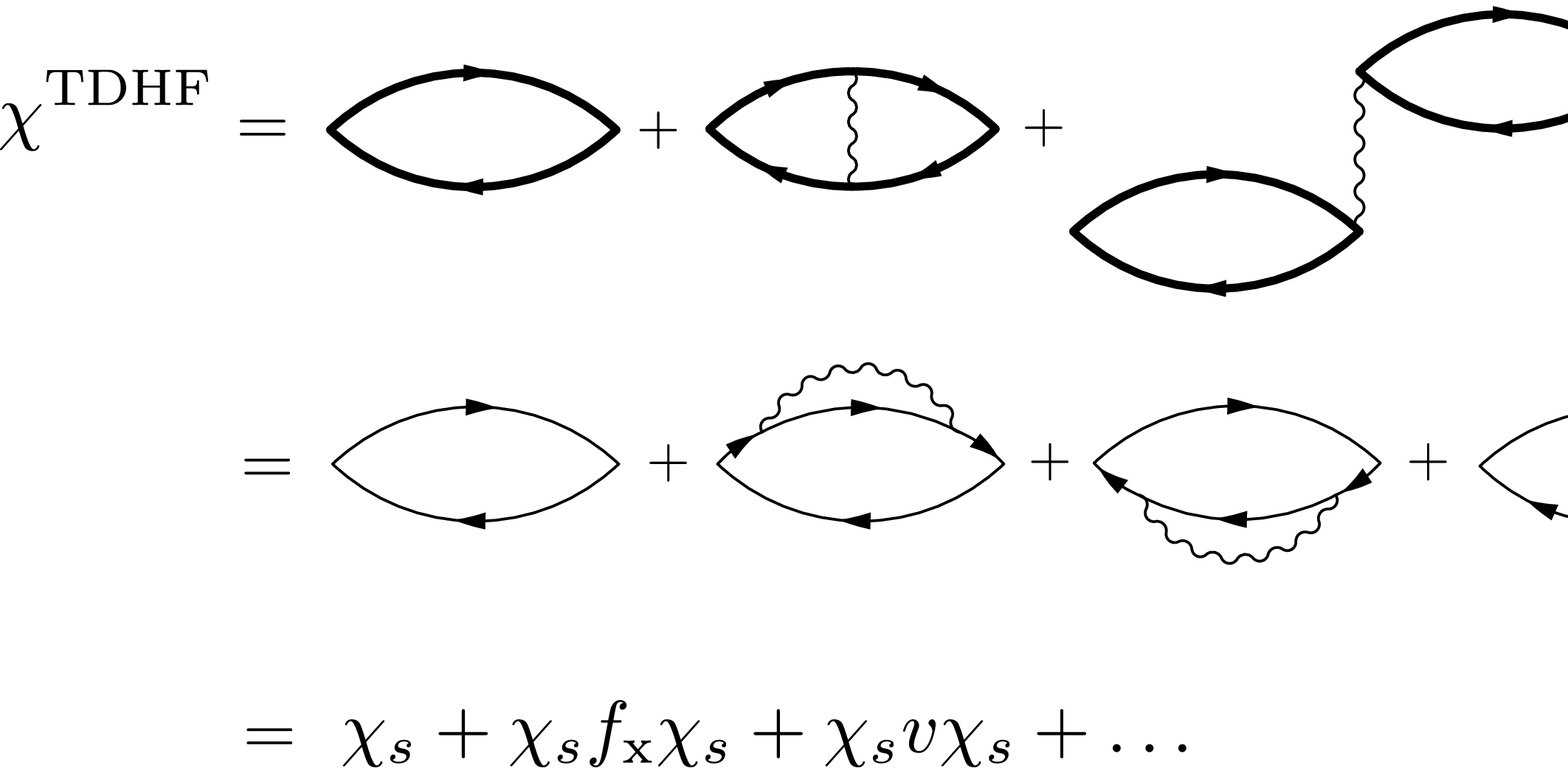}\\
\caption{The linear density response function in TDHF . The first row shows a diagrammatic expansion using the HF Green function. The second row shows an expansion in terms of the KS Green function. All diagrams up to first order are drawn and seen to be the same as the first order terms of $\chi$ in TDEXX.}
\label{diagram2}
\end{figure*}
The resulting equation for the EXX kernel $f_{\rm x}$ is represented diagrammatically
in Fig.~\ref{diagram}. Notice that the obtained $f_{\rm x}$ is often referred to as 
the kernel of the TDOEP (time-dependent optimized effective potential) method. It has been derived several times before by other people starting 
with Sharp and Horton\cite{sh} in the fifties and continuing with Talman and Shadwick\cite{ts} in the sixties. The resulting response function has been derived by Rajagopal,\cite{raja} Holas, Aravind and Singwi,\cite{holas} Lemmens,
Brosens and de Vrese,\cite{lem1,lem2} and more recently by Gross\cite{oepgross} to mention a few. Our way of deriving the same approximation has the advantage of demonstrating
the conserving properties of the approximation. These are properties also inherent in the original TDHF approximation. Below we will make further comparisons between TDHF and TDEXX, illustrating their similarities and differences. 
\subsection{TDEXX as compared to TDHF}
\label{hfvsxx}
The quantity to be determined in the TDHF equation is the three-point function $P~=~\delta G/\delta V_{\rm ext}$,~where $V_{\rm ext}$ is the external potential and $G$ is the HF 
Green function given by the solution of the Dyson equation, 
 \be
G=G_{\rm o}+G_{\rm o}\S^{\rm HF}[G]G,
\label{dyshf}
\ee
where $G_{\rm o}$ is the non-interacting Green function containing just the external (nuclear) potential and  $\Sigma^{\rm HF}=V_{\rm H}+\S^{\rm x}$ with $V_{\rm H}$ being the Hartree potential. By varying the Dyson equation with respect to the external potential we arrive at the equation for $P$, usually referred to as the TDHF equation
\bea
P(1,3;2)&\!=&\!G(1,2)G(2,3)\nonumber\\
&&\hspace{-10mm}+\int \!d4567\,G(1,4)G(5,3)\frac{\delta\S^{\rm HF}(4,5)}{\delta G(6,7)}P(6,7;2).
\label{bse}
\eea
The linear density response function can then be
obtained as $\chi^{\rm TDHF}(1,2)=-iP(1,1;2)$.
In Fig.~\ref{diagram2}, on the first row, $\chi^{\rm TDHF}$ is depicted diagrammatically with terms up to second order in the explicit dependence on the Coulomb interaction. There is, of course, also a dependence on $v$ through $G$, which is summed up to infinite order in the interaction strength through Eq.~(\ref{dyshf}). 

Instead of working with $G_{\rm o}$ as our zeroth order Green function we can choose to work with the KS Green function $G_s$. This Green function can also be found through a Dyson-like equation:
\be
G_s=G_{\rm o}+G_{\rm o}\{V_{\rm H}[G_s]+v_{\rm x}[G_s]\}G_s.
\label{dysvxc}
\ee
By inverting Eq.~(\ref{dyshf}) and Eq.~(\ref{dysvxc}) we find again the equation for $G$ but with $G_s$ as the zeroth order Green function:
\be
G=G_s+G_s\{\S^{\rm HF}[G]-V_{\rm H}[G_s]-v_{\rm x}[G_s]\}G.
\ee
Iterating to first order gives us
\be
G^{(1)}=G_s+G_s\{\S^{\rm x}[G_s]-v_{\rm x}[G_s]\}G_s.
\ee
The strictly first order diagrams (in terms of $G_s$) for $\chi^{\rm TDHF}$ can now be identified (see the second row in Fig.~\ref{diagram2}) and we see that they are identical to the first 
order terms of $\chi$ in the TDEXX approximation (see Eq.~(\ref{rpafxc}) and Fig.~\ref{diagram}).  Thus, we can conclude that, to first order in $v$, TDHF and TDEXX are the same. Notice that, in principle,  this conclusion is independent of the choice of $G_s$ and $v_{\rm x}$ as long as they are related via Eq. (\ref{dysvxc}). 
But, by using the self-consistent EXX potential the corresponding density is optimized to exactly reproduce the  HF density up to first order, thus minimizing the contribution of the approximate higher order terms in $\chi$. However, using any other reasonable $v_{\rm x}$ the corresponding $\chi$ in TDEXX is still expected to be rather close to the response function of TDHF. Indeed, as we shall see later (Sec. \ref{selfc}), using the EXX potential, the LDA potential or the exact XC potential leads to very similar response functions. The higher order terms of the TDEXX series can thus be interpreted as an approximation to the corresponding higher order diagrams of TDHF, where the frequency-independent four-point kernel of Eq. (\ref{bse}), i.e., $\delta\S^{\rm x}/\delta G$, is simulated by the frequency-dependent two-point kernel $f_{\rm x}$. The trick to approximate the beyond first order terms in 
the series of the TDHF response in terms of the zeroth and first order terms such that the whole series can be summed as a geometric one has been suggested before.\cite{holas} The full TDHF series can be written order by order as
\bea
\chi^{\rm TDHF}&=&\chi_{0}[1+\chi_{0}^{-1}\chi_{1}+\chi_{0}^{-1}\chi_{2}+\ldots]\approx\frac{\chi_{0}}{1-\chi_{0}^{-1}\chi_{1}}\nn\\
&=&\chi_{0}+\chi_{1}+\chi_{1}\chi_{0}^{-1}\chi_{1}+\ldots\nn
\eea
If $\chi_{0}=\chi_{s}$ this is just the TDEXX series.

As shown in the Appendix it is essential to derive the potential $v_{\rm x}$ from 
the LSS in order to have the $f$-sum rule obeyed. Self-consistency is, 
however, not necessary. Choosing any density there is a local one-body 
potential which, in a non-interacting system, generates that density as 
well as one-electron orbitals with corresponding eigenvalues. The LSS 
within EXX (Eq. (\ref{lsseq})) then gives a potential $v_{\rm x}$ which, together with 
these orbitals and eigenvalues, yields an $f_{\rm x}$ from Eq. (\ref{fxceq}). In this way, 
both  $v_{\rm x}$ and $f_{\rm x}$ are well defined functionals of the starting 
density. If the density is that given by EXX we have self-consistency 
and the resulting $f_{\rm x}$ is that presented here. If we instead start with a 
better density, much closer to the exact one, the corresponding eigenvalue 
differences will be much closer to real particle-hole excitation energies\cite{hvB} 
and we will obtain an $f_{\rm x}$ which will still obey the 
$f$-sum rule while giving rise to an optical spectrum with a continuum 
starting in almost the correct place - because the highest occupied exact 
DF eigenvalue equals the the negative of the ionization potential.\cite{ab} This topic is further discussed together with numerical results in Sec. \ref{selfc}.
\subsection{Numerical method}
The numerical implementation of the TDEXX approximation starts with a calculation 
of the self-consistent ground-state potential $v_{\rm x}$ from Eq.~(\ref{lsseq}). 
This potential determines the KS system from which $G_s$ and $\chi_s$ are calculated 
and inserted into Eq.~(\ref{fxceq}). The kernel is then obtained through multiplication by the inverse
of $\chi_s$ both from the left and the right. Finally, to obtain the full response function $\chi$ a RPA 
like equation needs to be solved, Eq.~(\ref{rpafxc}). Since $f_{\rm x}$ is frequency 
dependent all steps after the calculation of the potential need to be repeated at 
every frequency. Notice also that, due to the spherical symmetry of our studied 
systems, Eq.~(\ref{rpafxc}) separates into several decoupled equations, one for each 
angular momentum channel.  

We have chosen a basis set implementation using cubic splines as radial basis functions. 
This basis set has shown to be ideally suited for solving the LSS equation,\cite{hvB} 
an equation known to be numerically unstable when using other methods and basis 
functions.\cite{engeljiang,asympdiv,unique,unique2,soleng} We have here found that 
cubic splines also provide an efficient method for solving the equation for $f_{\rm x}$. 
In retrospect this might not be surprising since this equation has similarities to 
the LSS equation. 

A detailed description of the construction of our basis set can be found in 
Ref. \onlinecite{hvB} and a general discussion of B-splines in electronic structure 
calculations can be found in Refs. \onlinecite{martin,splines}. Here, we will only 
list the advantages of using these basis functions. 
1) To start with, they are local functions. This gives us a great amount of 
flexibility in choosing the distribution of splines. Where high accuracy is needed, 
like in our case close to the nucleus, the density of splines can be chosen 
arbitrarily high without loosing accuracy in other regions.
2) Another important property of the splines is that there is no risk of 
instabilities due to overcompletness because of the strong localization of the 
splines. 
3) Since every spline only overlaps with its three nearest neighbors all matrices 
will be band diagonal, reducing the amount of storage needed and allowing for 
the use of efficient diagonalization algorithms. 
4) Once a mesh distribution (e.g. a power law or an exponential distribution) 
and a maximum radius is set there is only one numerical parameter to vary, i.e., 
the number  $N$ of cubic splines.  
5) The basis set is complete. The results should thus converge to the exact results
as $N\ra\inf$. 
6) We have shown that the product of two orbitals can be re-expanded in the same 
basis set without increasing the number of basis functions. All two-particle functions, 
like for instance response functions, thus become matrices of the same order as one 
particle propagators.
7) A cubic spline is composed of cubic polynomials and hence all integrals can be
solved exactly, either analytically or, and actually faster by, using simple Gaussian quadrature. 

The work to calculate the density response function within just
the RPA (without exchange) is just about as extensive as that required
to obtain the response function from any so called adiabatic - or
frequency independent - approximation. The calculation of the exchange kernel $f_{\rm x}$ of the 
EXX involves sums over two continua for every frequency. An ordinary RPA calculation 
using $N$ basis functions requires for every frequency ${\rm const}*N^3$ operations because of the necessity to
invert an $N\times N$ matrix. Including also an exchange-correlation kernel
requires for each frequency a double sum over the continuum, i.e.,
${\rm const}*N^2$ operations plus two additional matrix inversions. This 
increases the prefactor of the $N^3$ dependence on the number $N$ of 
included splines. The prefector also depends heavily on the number 
of occupied states but, for large $N$, including the exchange kernel
does not constitute a qualitative difference compared to an ordinary
RPA calculation as far as the calculational effort is concerned.

To get an accurate description of both the occupied and the first few unoccupied
orbitals we used a cubic distribution of mesh points in all our calculations. 
The results were converged with $\sim 40$ splines for He and with $\sim 60$ 
splines for Ar. 
\section{Results}
\label{result}
In this Section we present our results on static polarizabilities, van der Waals 
coefficients and correlation energies for all spin-compensated spherical atoms up to Ar and some 
discrete excitation energies for Be and Ne. If not indicated all results are 
obtained with the self-consistent EXX Green function. The convergence criterion for the EXX potential was set to  $|n^{(k)}(\br)-n^{(k-1)}(\br)|\leq10^{-7}$. 
\begin{table}[t]
\caption{Static polarizabilities for some different atoms calculated in TDEXX, TDHF, RPA and from the KS system. (a.u.)}
\begin{ruledtabular}
\begin{tabular}{crrrrr}
Atom&\text{KS}&\mbox{RPA}&\mbox{TDEXX}&\text{TDHF}&\text{Litt.}\footnotemark[3]\\\hline
He&1.487&1.199&1.322&1.322\footnotemark[1]&1.38\\
Ne&2.838&2.234&2.372&2.377\footnotemark[1]&2.67\\
Ar&16.965&9.883&10.737&10.758\footnotemark[1]&11.08\\
Be&81.385&33.489&45.648&45.62\footnotemark[2]\,\,\,&37.8\,\,\,\\
Mg&140.26&60.262&81.658&81.60\footnotemark[2]\,\,\,&71.53\!\!
\label{statpol}
\end{tabular}
\end{ruledtabular}
\footnotetext[1]{From Ref. \onlinecite{poltdhf}}
\footnotetext[2]{From Ref. \onlinecite{poltdhfbemg}}
\footnotetext[3]{ From Ref. \onlinecite{polref}}
\end{table}
\subsection{Static polarizabilities}
The static polarizability is defined according to
\be
\alpha(0)=-\int z\chi(\br,\br',\w=0)z'd\br d\br'.
\label{pol}
\ee
For a system with spherical symmetry only the angular momentum channel $L=1$ of $\chi$ contributes. 
The small polarizabilities of the noble gas atoms as compared to the alkali earth atoms are due to the large gap in the excitation spectrum. On the contrary, Be and Mg have a near degeneracy in the HOMO-LUMO gap causing very large polarizabilities.  

In Table~\ref{statpol} we compare $\alpha(0)$ calculated in TDEXX with $\alpha(0)$ calculated in RPA, TDHF and from the KS response function, $\chi_s$. Calculating the static polarizability from the KS response function provides a reasonable estimate, albeit too large compared to the true static polarizability of noble gas atoms. In Be and Mg the error is much larger and leads to an overestimation by a factor of two. Including interaction effects at the level of RPA reduces the KS results for all atoms. However, the RPA polarizabilities are consistently too low as compared to more accurate values. Introducing exchange effects at the level of TDEXX increases $\alpha(0)$ again leading to an appreciable improvement for the noble gas atoms but the error for Be and Mg remain roughly the same but with a different sign. 

The TDEXX polarizabilities are seen to be very close to those of TDHF. This is not surprising as TDDFT within the EXX can be considered as a variational solution to the integral equation of TDHF theory. For a two-electron system like He TDEXX is actually identical to TDHF. 
\begin{table}[t]
\caption{van der Waals coefficients calculated in different approximations.  (a.u.)}
\begin{ruledtabular}
\begin{tabular}{crrrrr}
Atom &KS&RPA&TDEXX& TDHF&Litt. \\
\hline
He&1.664&1.171&1.375&1.375\,\,\, &1.458\footnotemark[3]\\
Ne&7.492&5.003&5.506&5.524\footnotemark[1] & 6.383\footnotemark[3]\\
Ar&128\,\,\,\,\,\,\,\,\,\,\,&54.23\,\,\,&61.88\,\,\,&61.88\footnotemark[1]\,\,\, &64.30\footnotemark[3]\,\,\,\\
Be&660\,\,\,\,\,\,\,\,\,\,\,&179\,\,\,\,\,\,\,\,\,\,\,&283\,\,\,\,\,\,\,\,\,\,\,&284\footnotemark[2]\,\,\,\,\,\,\,\,\,\,\,&214\footnotemark[4]\,\,\,\,\,\,\,\,\,\,\\
Mg&1723\,\,\,\,\,\,\,\,\,\,\,&482\,\,\,\,\,\,\,\,\,\,\,& 765\,\,\,\,\,\,\,\,\,\,\,&758\footnotemark[5]\,\,\,\,\,\,\,\,\,\,\,&627\footnotemark[4]\,\,\,\,\,\,\,\,
\label{vander}
\end{tabular}
\end{ruledtabular}
\footnotetext[1]{From Ref. \onlinecite{shh2006}.}
\footnotetext[2]{From Ref. \onlinecite{tdhfbe}.}
\footnotetext[3]{From Ref. \onlinecite{dosd}.}
\footnotetext[4]{From Ref. \onlinecite{vdwbemg}.}
\footnotetext[5]{From Ref. \onlinecite{vdwmg}.}
\end{table}
\subsection{van der Waals coefficients}
The van der Waals coefficient, or $C_6$-coefficient, between ion $A$ and $B$ is  given by the formula
\be
C_6=\frac{3}{\pi}\int_0^\infty \alpha_A(i\w)\alpha_B(i\w)d\w,
\ee
where $\alpha_A(i\w)$ is the dynamic polarizability of ion $A$ calculated at imaginary frequencies. 
In Table~\ref{vander} the $C_6$-coefficients in the TDEXX approximation are presented and compared to the values of the KS system, values calculated in the RPA and the TDHF approximation, as well as accurate values found in the literature.  Although TDEXX significantly improve over the RPA for He, Ne and Ar the $C_6$ coefficients remain too small. The results for Mg and Be in TDEXX are too large and give no improvement over RPA in an absolute sense. The TDEXX values is again seen to be in good agreement with the full TDHF results. This was also noted in Ref. \onlinecite{shh2006} for He, Ne and Ar.
\begin{table}[b]
\caption{Correlation energies calculated within different approximations and compared to accurate CI calculations. The correlation energy is here defined as the total energy minus the HF energy. In the last column also the Hartree-Fock total energies are tabulated.  (a.u.)}
\begin{ruledtabular}
\begin{tabular}{crrrrr}
Atom &TDEXX&RPA &MP2\footnotemark[1] &CI\footnotemark[2]&HF\footnotemark[2]\\\hline
He&0.044&0.083&0.047&0.0420& 2.8617\\
Ne&0.389&0.596&0.480&0.3905&128.5471\\
Ar&0.721&1.091&0.844&0.7225&526.8175\\
Be&0.102&0.181&0.124&0.0943&14.5730\\
Mg&0.445&0.681&0.514&0.4383&199.6146\!\!
\label{toten}
\end{tabular}
\end{ruledtabular}
\footnotetext[1]{From Ref. \onlinecite{engeljiang}.}
\footnotetext[2]{From Ref. \onlinecite{CI}.}
\end{table}
\subsection{Correlation energies}
\begin{figure}[t]
\includegraphics[width=8.5cm, clip=true]{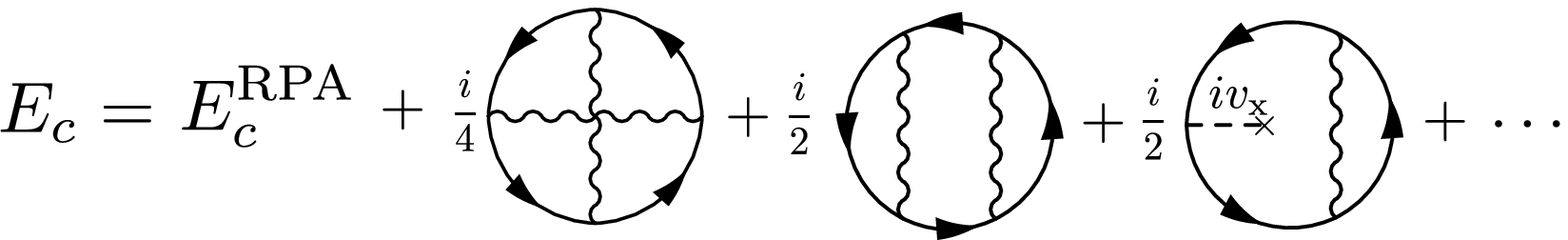}\\
\caption{Diagrams contained in the correlation energy functional, Eq.~(\ref{corr}), with $f_\xc=f_{\rm x}.$}
\label{ediag}
\end{figure}

Using the standard trick based on the Hellman-Feynman theorem to integrate the interaction energy with respect to the strength of the Coulomb interaction the correlation energy becomes
\be
E_c=-\int_0^1 d\lambda \int_0^\infty \frac{d\omega}{2\pi}\,\,{\rm Tr}\{v[\chi^{\lambda}(i\omega)-\chi_s(i\omega)]\}.
\label{corr}
\ee
In Eq.~(\ref{corr}) we have use the short hand notation $\Tr fg=\int d{\bf r}d{\bf r'} f({\bf r},{\bf r'})g({\bf r'},{\bf r})$ for any two-point functions $f$ and $g$, and defined the response function
$$
\chi^{\lambda}=\frac{\chi_s}{1-(\lambda v+f^{\lambda}_{\rm xc})\chi_s},
$$
where $f^{\lambda}_{\rm xc}$ is the XC kernel of a system of electrons interacting through the 
rescaled Coulomb potential $\l v$. From this expression the simplest approximation is obtained by setting $f_{\rm xc}=0$, leading to the formula for the RPA correlation energy. A fully self-consistent calculation of this approximation was performed in Ref. \onlinecite{hvB} for all spin-compensated spherical atoms up to Ar. With $f_\xc=f_{\rm x}$ a cancellation between Hartree and exchange terms is expected to occur, improving the largely overestimated RPA values. We have here, for the first time, performed such a calculation. The results are presented in Table~\ref{toten} and compared with those in the RPA and the MP2 approximation\cite{engeljiang} as well as results from accurate CI calculations.\cite{CI} As expected, the TDEXX results are very accurate. A diagrammatic analysis (see Fig.~(\ref{ediag})) shows that with this kernel the correlation energy will, apart from the RPA or bubble series of diagrams, also contain the important second order exchange diagram (included in the MP2 approximation) as well as an infinite series of terms simulating the higher order exchange diagrams.
\subsection{$f$-sum rule and EXX kernel}
In the Appendix we prove that the $f$-sum rule, a consequence of particle conservation, is valid in the TDEXX approximation by showing that the coefficient of the $1/\w^2$ term in the large $\w$ expansion of $\chi$ is the same as in the expansion of $\chi_s$. Thus, by expressing the dynamical polarizability, Eq.~(\ref{pol}), as
\be
\alpha(i\w)=\sum_q\frac{f_q}{\w^2+\w^2_q},
\label{fqq}
\ee
where  $q=(k,\mu)$ is a particle-hole index, $\omega_q$ is an excitation energy and $f_q$ is the corresponding oscillator strength, the sum over all oscillator strengths must equal the number of particles
$$
\sum_qf_q=N.
$$
As a test of our numerical accuracy this result was checked. We multiplied $\alpha(i\w)$ by $\w^2$ and studied the large $\w$-values of this function. Indeed, within high accuracy ($\leqslant 10^{-4}$), the results converged to $N$ for all atoms. 
\begin{figure}[t]
\includegraphics[width=8.cm, clip=true]{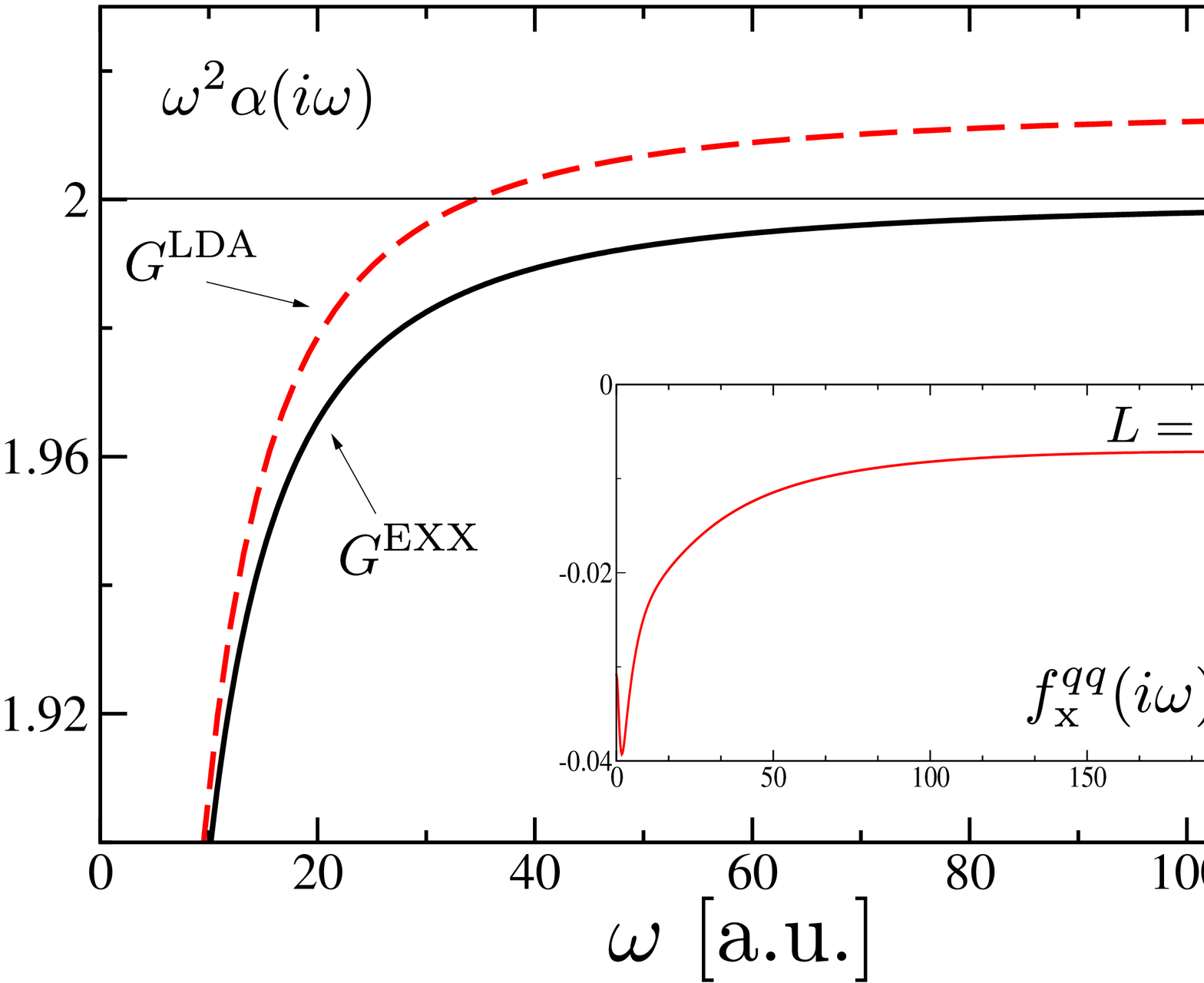}\\
\caption{The main figure shows $\w^2\alpha(i\w)$ for He calculated at different Green functions. The inset shows $f^{qq}_{\rm x}$ with $L=1$ and $q=2s\rightarrow2p$ for Be. The large $\w$-behavior in both plots clearly indicates that the $f$-sum rule is obeyed in our calculations.}
\label{fsum}
\end{figure}

In Fig.~\ref{fsum} we have plotted $\w^2\alpha(i\w)$ for He. The kernel $f_{\rm x}$ was calculated using either the Green function $G_s$ and the exchange potential $v_{\rm x}$ of the EXX or the same quantities within LDA. The $f$-sum rule is not expected to be obeyed in the latter case. And this is, indeed, what is observed. It should be noticed, however, that the violation is minor ($ \sim 0.8\%$).

\begin{figure}[b]
\includegraphics[width=8.cm, clip=true]{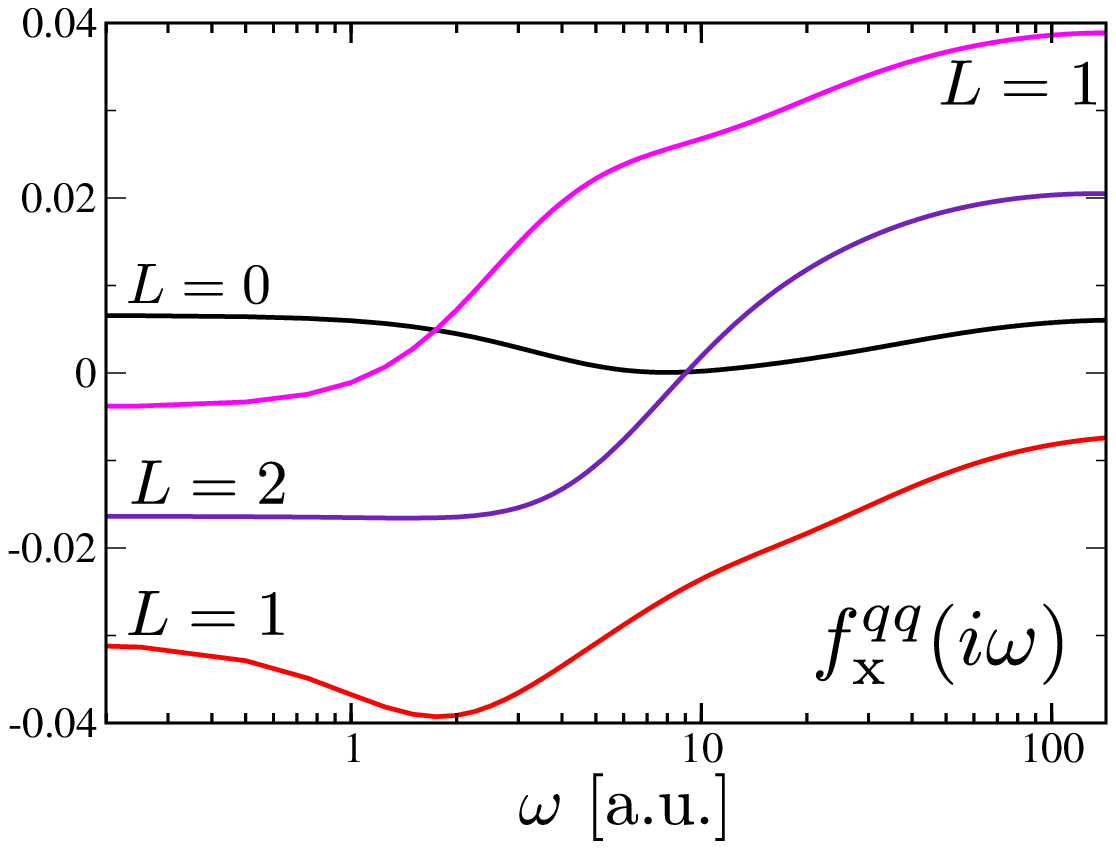}\\
\caption{The quantity $f^{qq}_{\rm x}$ for Be is plotted for different $L$ at imaginary frequencies. There are two curves with $L=1$. In the lower, $q$ corresponds to the $2s\rightarrow2p$ transition and in the upper to the $1s\rightarrow2p$ transition.}
\label{ploim}
\end{figure}

The quantity $\chi_sf_{\rm x}\chi_s$ must decay as $1/\w^4$ (see the Appendix). As a consequence, $f_{\rm x}$ cannot diverge (as it does, e. g., in a model system, see Ref. \onlinecite{argunn}) and must approach a constant as $\w\rightarrow\infty$. Another way to check the $f$-sum rule and to test our calculations is thus to study the large $\w$-behavior of $f_{\rm x}$. To do this we first notice that the kernel only appears in the form of matrix elements of the $F_q$-functions: 
$$
f^{qq'}_{\rm x}(\omega)=\int F_q(\br)f_{\rm xc}(\br,\br',\omega)F_{q'}(\br')d\br d\br',
$$
where $F_q$ is a KS excitation function, i. e., a product of the occupied
KS orbital $\varphi_k$ and the unoccupied KS orbital $\varphi_{\mu}$. 
Now, since the excitation functions alone integrates to zero the kernel is unique only up to the addition of  two arbitrary functions, $g_1(\omega,\br)$ and $g_2(\omega,\br')$.  The quantity $f^{qq'}_{\rm x}$ is unique though and in Fig.~\ref{ploim} we have plotted this quantity for Be at imaginary frequencies and different $L$. The frequency dependence is approximately constant for low frequencies but becomes more pronounced for higher frequencies. This justifies the use of the adiabatic approximation for low energies. At large $\w$ every matrix elements $f^{qq'}_{\rm x}(i\w)$ is seen to approach a constant. This is again demonstrated for one matrix element in the inset of Fig.~\ref{fsum}. 
\subsection{Excitation energies}
The poles of the exact linear density response function correspond to the particle conserving excitation energies of the system. 
Thus, any approximate $\chi$ yields an approximate set of excitation energies. $\chi_s$ contains the excitation energies of the KS system and in many cases they give a good approximation to the discrete part of the spectrum. Nevertheless, $\chi_s$ is a non-interacting response function and as such lacks many features of the full many body $\chi$. 

With $f_\xc=0$, Eq.~(\ref{rpafxc}) can be rewritten as an eigenvalue problem where the eigenvalues correspond to the poles of $\chi$. Due to the frequency dependence of the XC kernel of the EXX this technique to obtain the excitation energies cannot be applied in a straight-forward way.
In the present work we have chosen to work with localized basis functions
which causes ${\rm Im}\chi$ as a function of frequency to consist of a series of
sharp delta functions. And this behavior is independent of whether the frequency
lies in the continuum or in the discrete part of the spectrum.
In order to obtain something which can be plotted we arbitrarily add a small
positve imaginay part to the real frequency and obtain the discrete excitation
energies from the positions of the resulting huge peaks in ${\rm Im}\chi$.

In Table~\ref{Beexc} we present the first few discrete excitation energies of Be and Ne and compare them to experimental values and to the ones obtained in RPA, TDHF and the KS system (here, meaning the differences between the KS one-electron eigenvalues). In the case of Be the KS values are too low. The RPA improves over the KS results but proceeding to the TDEXX makes them, somewhat worse. 
It is interesting to observe how close the results of the TDEXX are to those of the TDHF. In the case of Ne we have observed a qualitatively different behavior. The KS eigenvalue differences are larger than the experimental values.
As in the case of Be, the RPA tends to increase the  excitation energies yielding, for Ne, an even larger discrepancy.
The inclusion of exchange effects does not lower the RPA results, as in the Be case, but rather increase them.
Although TDEXX and TDHF push the RPA excitation energies in the same direction the actual values are not as close as for Be. 

The oscillator strengths for the discrete excitation energies can be extracted from the height of the peaks
in the optical spectrum. As an example, we obtain 1.379 for the lowest transition of Be. This can be compared to the TDHF value of 1.378.\cite{tdhfbe} 

\subsection{Sensitivity to the ground state KS approximation}
\label{selfc}
As noticed by others,\cite{pgb} it is crucial to have good KS transitions as a starting point when calculating the discrete excitation energies. With the exact KS potential\cite{umr} for Ne the $2p\rightarrow3s$ transition in the KS system becomes 0.612, an already very good approximation to the true value (0.619). The transition frequencies we have obtained are thus expected to be improved by using a ground-state potential better than that of EXX. When we improve the  ground-state density and the corresponding orbitals and eigenvalues we still calulate $v_{\rm x}$ from the LSS in order not to violate the sum-rule as discussed in Sec. \ref{hfvsxx}. 

We have used the exact densities of Umrigar et. al.\cite{umr} and found the static polarizabilities 1.35 and 40.5 for He and Be, respectively. There is, thus, a significant improvement, particularly in the case of Be. This shows that the largest error is actually caused by a poor description of the ground state within the EXX. With the very accurate densities by Umrigar et. al. the transition $2s\rightarrow2p$ in Be becomes 0.182 and the $2p\rightarrow3s$ transition in Ne becomes 0.631. The transition frequencies are thus also largely improved by using an accurate KS ground state.  

By also using the accurate XC potentials by Umrigar et. al. in Eq. (\ref{fxceq}) for generating $f_{\rm x}$ we violate the $f$-sum rule but, in this way, as discussed in Sec. \ref{hfvsxx}, we obtain a response function which to first order in the Coulomb interaction is identical to that of TDHF. We, therefore, expect the results to revert to our original results from TDEXX which are close to those of TDHF. Indeed, we now obtain the static polarizabilities 1.323 and 45.76 for He and Be, compared to our previous results 1.322 and 45.65 respectively. 
Using instead the LDA potential for generating $f_{\rm x}$ we obtain 1.343 and 45.23 for the same quantities, again demonstrating the closeness to the TDHF.
\begin{table}[t]
\caption{The first few discrete excitation energies for Be and Ne in TDEXX compared to experimental, TDHF, RPA and KS transitions.}
\begin{ruledtabular}
\begin{tabular}{lcccccl}
&Transition&KS&RPA&TDEXX&TDHF\footnotemark[1]&Exp.\footnotemark[2] \\
\hline
Be&&&&\\
&2s$\rightarrow$2p& 0.1312 & 0.2032  & 0.1764 &0.1764 & 0.1940 \\
&2s$\rightarrow$3p& 0.2412 & 0.2547 & 0.2470 &0.2471 & 0.2742  \\
&2s$\rightarrow$4p& 0.2731 &   0.2777& 0.2749 &0.2750 & 0.3063 \\
&2s$\rightarrow$5p&  0.2868   &  0.2889 & 0.2877&0.2878 &  0.3195         \\
Ne&&&&\\
&2p$\rightarrow$3s& 0.6585  & 0.6675 & 0.6803 &  0.6739  &0.6190 \\
&2p$\rightarrow$4s& 0.7793 &  0.7812  &  0.7827  &  0.7818  & 0.7268\\
&2p$\rightarrow$5s& 0.8134  &0.8141& 0.8147&  0.8139 &0.7593
\label{Beexc}
\end{tabular}
\end{ruledtabular}
\footnotetext[1]{From Refs. \onlinecite{tdhfbe,tdhfne}.}
\footnotetext[2]{Adopted from Refs. \onlinecite{tdhfbe,tdhfne}.}
\end{table}
\section{Conclusions and Outlook}
\label{conclu}
In this paper we have calculated the linear density response function of the 
TDEXX approximation for all spin-compensated spherical atoms up to Ar. For the properties 
studied in this work, 
i. e., static polarizabilities, van der Waals coefficients and the low-lying excitation energies, the results 
show that TDEXX is a good approximation to TDHF. 

TDEXX only takes into account exchange effects in the response of the system to external perturbations. 
For noble  gas atoms  the static polarizabilities and van der Waals coefficients still turn out to be in rather good agreement with the experimental values. The relative error is about $5\%$ for He, $13\%$ for Ne and $3\%$ for Ar. On the other hand for alkali earth atoms the results are not as satisfactory, the relative error being  around $26\%$ for Be and $18\%$ for Mg. For these systems it is thus necessary to include correlation effects in order to get an accurate description of the above properties. The low-lying excitation energies are accurate to within $10\%$ for both Be and Ne. We show here, however, that significantly better excitation energies as well as static polarizabilities can be obtained by using a more accurate exchange correlation potential for the ground state.

We have also calculated the total energies from the Hellman Feynman theorem applied to the strength of the Coulomb interaction. The results are in excellent agreement with accurate CI calculations. This is probably due to the fact that the fluctuation-dissipation formula at the level of TDEXX accounts for 
several correlation diagrams as, for instance, the important second order exchange diagram.
 
Finally, we have examined the behavior of kernel $f_{\rm x}$ along the imaginary frequency axis and found it to be both weakly and slowly dependent on $\w$. It is, however, not without structure indicating a more complex behavior on the real axis as compared to that resulting from simple poles. Indeed, we have seen that the kernel has both single and double poles on the real axis. The weak frequency dependence at the imaginary axis suggests that the adiabatic, i.e., frequency independent approximations to the EXX approximation, might not be such a bad idea at least not as far as total energies are concerned. The experience from the electron gas, however, strongly contradicts this conjecture.\cite{almhind}

In most cases of approximations within DFT and TDDFT it is very 
difficult to see through what kind of physical processes are actually
incorporated into that particular approximation. In our case, basing
our approximations on $\Phi$-derivable theories within MBPT, we can say
with confidence that a description of double excitations is way above
the EXX. Such a description would require an $f_{\rm xc}$ based at least on 
the time-dependent GW approximation which, by the way, probably would
yield much better van der Waals coefficients. But again we refer this discussion to a future publication.
\begin{acknowledgments}
This work was started in collaboration with S. Kurth when he was still in Lund and we thank S. Kurth for sharing his experience from this problem. We also thank C.-O. Almbladh for useful discussions. 
This work was supported by the European Community Sixth Framework Network of Excellence 
NANOQUANTA (NMP4-CT-2004-500198).
\end{acknowledgments}
\appendix
\section{$f$-sum rule}
In Ref. \onlinecite{tddftvar} we demonstrated that every $f_\xc$ of TDDFT obtained from the variational formulation of that work obeys particle conservation, which amounts to the $f$-sum rule in the linear limit.

In this Appendix we will demonstrate how the $f$-sum rule explicitly comes out of the construction of the kernel $f_{\rm x}$ of the EXX approximation within TDDFT. Among other virtues, this detailed derivation demonstrates the crucial importance of using the LSS equation when constructing the kernel $f_{\rm x}$ in order to have the sum rule fulfilled. From Eq.~(\ref{rpafxc}) we see that the total density response function $\chi$ can be written as
$$
\chi=\left[1-(v+f_{\rm x})\chi_s\right]^{-1}\chi_s
$$
Since the ground-state KS theory gives the correct density it follows that $\chi_s$ obeys the $f$-sum rule. Given the Lehmann representations for the response functions $\chi$ and $\chi_s$ this ensures that the coefficient of the $1/\w^2$ term of the large $\w$ expansion of $\chi_s$ has the correct value. If the full response function $\chi$ is also to obey the $f$-sum rule there must, obviously, be no contribution to the $1/\w^2$ coefficient from the denominator which thus must tend to unity at large $\w$. A sufficient condition for this is that $f_{\rm x}\chi_s$ vanishes as $1/\w^2$ in this limit. From Fig.~\ref{diagram} we see that the quantity which is actually calculated from the diagrams is $\chi_sf_{\rm x}\chi_s$ which thus should decay as $1/\w^4$ given the $1/\w^2$ of dependence of $\chi_s$ in this limit. In order to see that this is indeed the case we explicitly examine the large frequency behavior of the different contributions to $\chi_sf_{\rm x}\chi_s$ . As seen from Fig.~\ref{diagram} this quantity has contributions from five Feynman diagrams of which those proportional to the exchange potential $v_{\rm x}$ are most easily combined with one each of the two self-energy diagrams. The contribution from one of the diagrams with self-energy insertions is
\bea
S_1(\br,\br';\w)=2i\int\frac{d\w'}{2\pi} G_s(\br_1,\br;\w')\ G_s(\br',\br_2;\w')\nonumber\\
\times G_s(\br,\br';\w+\w')\Delta(\br_1,\br_2)\nonumber
\eea
where $\Delta=v(\br,\br')\sum_kn_k\vf_k(\br)\vf_k^*(\br')+v_{\rm x}(\br)\delta(\br-\br')$ and $n_k$ is 1 for occupied states and zero otherwise. Due to time-reversal symmetry the non-interacting Green functions are symmetric in there spatial arguments. It is then easily seen that the contributions from the remaining two diagrams with self energy insertions is obtained by adding to $S_1(\w)$ above the result $S_1(-\w)$. Consequently, the sum of all diagrams with self energy insertions is an even function of $\w$. Carrying out the frequency integrations we obtain
\bea
S_1(\br,\br';z)=\sum_{k_1k_2k_3}\bra k_2|\Delta|k_3\ket\vf_{k_1}(\br)\vf_{k_1}^*(\br')\vf_{k_2}(\br)\vf_{k_3}^*(\br')\nonumber\\
\times\frac{2}{\ve_{k_3}-\ve_{k_2}}\left\{\frac{n_{k_1}-n_{k_3}}{z+\ve_{k_1}-\ve_{k_3}}-\frac{n_{k_1}-n_{k_2}}{z+\ve_{k_1}-\ve_{k_2}}\right\}\nonumber
\eea
Here, the functions $\vf_k(\br)$ are the KS orbitals with eigenvalue $\ve_k$ and we have switched from time-ordered quantities to retarded ones by everywhere replacing $\w-i\delta$ by $\w+i\delta$. The quantity $z$ is a complex frequency in the upper half plane.

When $z\rightarrow\infty$ the contribution appears to behave as $1/z$ but we remind the reader that we should add a similar expression with $z$ replaced by $-z$. Then, the leading order becomes $A/z^2$ with the coefficient
\bea
A=4\sum_{k_1k_2k_3}\bra k_2|\Delta|k_3\ket\vf_{k_1}(\br)\vf_{k_1}^*(\br')\vf_{k_2}(\br)\vf_{k_3}^*(\br')\nonumber\\
\times\left\{\frac{(n_{k_1}-n_{k_3})(\ve_{k_3}-\ve_{k_1})}{\ve_{k_3}-\ve_{k_2}}-\frac{(n_{k_1}-n_{k_2})(\ve_{k_2}-\ve_{k_1})}{\ve_{k_3}-\ve_{k_2}}\right\},\nonumber
\eea 
where we have multiplied by two to account for the remaining two self-energy like diagrams $(S_1(-\w))$. We can regroup the terms and write $A=A_1+A_2$ where
\bea
A_1=4\sum_{k_1k_2k_3}\bra k_2|\Delta|k_3\ket\vf_{k_1}(\br)\vf_{k_1}^*(\br')\vf_{k_2}(\br)\vf_{k_3}^*(\br')\nonumber\\
\times(n_{k_1}-n_{k_3})\nonumber\\
A_2=4\sum_{k_1k_2k_3}\bra k_2|\Delta|k_3\ket\vf_{k_1}(\br)\vf_{k_1}^*(\br')\vf_{k_2}(\br)\vf_{k_3}^*(\br')\nonumber\\
\times(n_{k_3}-n_{k_2})\frac{\ve_{k_1}-\ve_{k_2}}{\ve_{k_3}-\ve_{k_2}}\nonumber
\eea
Let us now define the one-particle density matrix by
$$
n(\br,\br')=2\sum_kn_k\vf_k(\br)\vf_k^*(\br')
$$
and use the completeness of the KS orbitals. We obtain
\bea
A_1=\delta(\br-\br')\int d^3\br_3v(\br-\br_3)|n(\br_3,\br)|^2\nonumber\\-v(\br-\br')|n(\br,\br')|^2
\label{A1}
\eea
and see that the diagrams containing $v_{\rm x}$ do not contribute to leading order in the large frequency limit.

In order to manipulate the $A_2$ coefficient we use the fact that the KS orbitals obey the KS equation
$$
\left\{-\frac{1}{2}\nabla^2+V(\br)\right\}\vf_k(\br)=\ve_k\vf_k(\br)
$$
where $V(\br)$ is the full KS potential. Because of the difference $\ve_{k_1}-\ve_{k_2}$, the terms involving the potential 
will vanish and we obtain
\bea
A_2=2\sum_{k_1k_2k_3}\bra k_2|\Delta|k_3\ket\vf_{k_1}^*(\br')\vf_{k_3}^*(\br')\frac{n_{k_3}-n_{k_2}}{\ve_{k_3}-\ve_{k_2}}\nonumber\\
\times\left\{\vf_{k_1}(\br)\nabla^2\vf_{k_2}(\br)-\vf_{k_2}(\br)\nabla^2\vf_{k_1}(\br)\right\}\nonumber
\eea
Now, using
$$
\vf_1\nabla^2\vf_2-\vf_2\nabla^2\vf_1=2\nabla(\vf_1\nabla\vf_2)-\nabla^2(\vf_1\vf_2)
$$
and the completeness we can write $A_2$ as
$$
A_2=2\nabla\left[\delta(\br-\br')\nabla f(\br)\right]-2\nabla^2\left[\delta(\br-\br')f(\br)\right]
$$
where
$$
f(\br)=\sum_{k_2k_3}\bra k_2|\Delta|k_3\ket\vf_{k_2}(\br)\vf_{k_3}^*(\br)\frac{n_{k_3}-n_{k_2}}{\ve_{k_3}-\ve_{k_2}}
$$
This is because, by symmetry,
$$
\nabla f(\br)=2\sum_{k_2k_3}\bra k_2|\Delta|k_3\ket\vf_{k_3}^*(\br)\nabla\vf_{k_2}(\br)\frac{n_{k_3}-n_{k_2}}{\ve_{k_3}-\ve_{k_2}}
$$
But, $f(\br)=0$ is the LSS equation defining $v_{\rm x}$ and we have shown that $A_2=0$.

Let us now study the remaining fifth diagram - the vertex diagram - in the high frequency limit. The contribution $R_V$ from this diagram is
\bea
R_V(\br,\br';\w)=2\int d^3\br_1d^3\br_2\int\frac{d\w_1}{2\pi}\int\frac{d\w_2}{2\pi}\nonumber\\
\times G_s(\br,\br_1;\w_1+\w)G_s(\br_1,\br';\w_2+\w)\nonumber\\
\times v(\br_1,\br_2)G_s(\br',\br_2;\w_2)G_s(\br_2,r;\w_1)\nonumber
\eea
Carrying out the frequency integrals and converting to the retarded propagator in the upper half plane ($\w\rightarrow z$ with ${\rm Im}z > 0$) gives
\bea
R_V(\br,\br';\w)=-2\sum_{k_1k_2}\sum_{k'_1k'_2}\vf_{k_1}(\br)\vf_{k_2}^*(\br)\vf_{k'_1}^*(\br')\vf_{k'_2}(\br')\nonumber\\
\times\bra k_1k'_2|v|k'_1k_2\ket\frac{(n_{k_1}-n_{k_2})(n_{k'_1}-n_{k'_2})}{(z+\ve_{k_2}-\ve_{k_1})(z+\ve_{k'_2}-\ve_{k'_1})}\nonumber
\eea
Here, the standard Coulomb integral is given by
\bea
\bra k_1k'_2|v|k'_1k_2\ket=\int d^3rd^3r'\vf^*_{k_1}(\br)\vf_{k'_2}^*(\br')v(\br-\br')\nonumber\\
\times\vf_{k'_1}(\br)\vf_{k_2}(\br')\nonumber
\eea
In the high frequency limit, to leading order, this becomes $B/\w^2$ where the coefficient $B$ is given by
\bea
B=-2\sum_{k_1k_2}\sum_{k'_1k'_2}\vf_{k_1}(\br)\vf_{k_2}^*(\br)\vf_{k'_1}^*(\br')\vf_{k'_2}(\br')\nonumber\\
\times\bra k_1k'_2|v|k'_1k_2\ket(n_{k_1}-n_{k_2})(n_{k'_1}-n_{k'_2})\nonumber
\eea
Using again the completeness and the definition of the density matrix $n(\br,\br')$ we obtain
\bea
B=-\delta(\br-\br')\int d^3\br_3v(\br-\br_3)|n(\br_3,\br)|^2\nonumber\\
+v(\br-\br')|n(\br,\br')|^2\nonumber
\eea
This is just the coefficient $A$, Eq.~(\ref{A1}), with the opposite sign. Consequently, the $1/z^2$ contribution from the vertex diagram exactly cancels the same term from the self-energy contributions meaning that $\chi_sf_{\rm x}\chi_s$ decays as $1/z^4$ at large frequencies. And this provides an explicit proof of the $f$-sum rule in the EXX approximation.

Finally we note that this result means that the exchange kernel should have a very weak dependence on frequency at large frequencies.

\end{document}